\def\aa{{A\&A}}

\def\apj{{ApJ}}
\def\apjs{{ApJS}}

\def\mnras{{MNRAS}}

\documentclass[cup5b]{caps}
\usepackage{graphicx}
\usepackage{amssymb}
\usepackage{ociwsymp4}   %Use this for invited papers.

\HeadText{B. E. J. Pagel}  %Enter author name; if three authors, list all three;
\begin{document}

\pagenumbering{arabic}

%Author names should be in captital letters
\author[]{B. E. J. PAGEL\\ Astronomy Centre, University of Sussex, UK}  

%Example for multiple authors:
%
%\author[]{D. C. BACKER$^{1}$, A. H. JAFFE$^{2}$, and A. N. LOMMEN$^{3}$
%\\
%(1) University of California, Berkeley, CA, USA\\
%(2) Imperial College, London, England\\
%(3) Sterrenkundig Instituut ``Anton Pannekoek'', Amsterdam, The Netherlands}
%

\chapter{ Summary Talk}

\begin{abstract}
Highlights of the meeting include new insights into {\em r} and {\em s-}processes 
and an explosion of interesting abundance data on stars in the Galactic halo 
and in dwarf spheroidal galaxies. I include in this summary a few suggestions on 
the chemical evolution of the Thick and Thin disks, and on the present and past 
distribution of baryons and metals in the universe.  

\end{abstract}

\section{Introduction}

It has been a great pleasure and honour for me to be able to attend this 
fascinating Symposium and I should like to express my warm thanks to Andy 
McWilliam and Michael Rauch for inviting me. Clearly the Carnegie Observatories 
(or Mount Wilson as I still like to think of them) are carrying their illustrious 
traditions forward into their second century of existence.  

\section{History}

George Preston recalled some highlights associated with the names of G.E. Hale, A.S. 
King, H.N. Russell, Horace Babcock, Lawrence Aller, Paul Merrill et al. I was 
intrigued to hear that after the discovery of stellar technetium Merrill began to 
pay much more attention to unconventional new ideas; I think he would have 
enjoyed this conference! 

Margaret Burbidge described the steps leading up to current ideas on nucleosynthesis, 
beginning with Hoyle's pioneering work in 1946 through the peaks in the 
standard abundance distribution and her work with Geoff on the barium star 
HD 46407 to B$^ 2$FH and subsequent developments. I remember visiting Margaret 
and Geoff in their small apartment in Botolph Lane, Cambridge, all covered in 
tracings, in 1954.     

\section{The overall picture} 

George Wallerstein raised some of the questions that form a backdrop to much of 
our proceedings, notably the issue of galaxy formation by monolithic collapse 
versus a series of mergers and the extent to which the Galactic halo could have been 
formed from dwarf spheroidals.  Abundance characteristics like the $\alpha$/Fe 
ratio preclude such a possibility for the classical globular clusters, but some field 
stars, especially with retrograde orbits, might originate in captured dwarfs.  

\section{Nucleosynthetic yields} 
\subsection{Massive stars} 
Yields from massive stars were discussed by Dave Arnett, Claudia Travaglio and 
Ken Nomoto.  
There are basically two kinds of process: hydrostatic 
burning (hydrogen, helium, carbon and neon burning and {\em s}-process) and 
explosive synthesis 
(oxygen and silicon burning and the {\em r} and {\em rp}-processes). 
Laser plasma experiments 
have led to improved knowledge of opacities and equations of state, but the 
structure of supernova remnants like Cas A shows up the limitations of 
one-dimensional models; 2D and 3D models are under development and help to 
`educate' the 1D models, e.g. by discovering instabilities, 
but the explosion mechanism is still uncertain. 

The abundance peculiarities of extremely metal-deficient stars 
(notably Christlieb's star HE 0107 $-$5240 with [Fe/H] = $-$5.3), 
include excesses (relative to iron) of C, N, O, $\alpha$-elements, Zn and Co 
and deficiencies of Cr and Mn. Ken Nomoto was able to explain these abundance 
patterns from nucleosynthesis by Pop III supernovae with masses anywhere between 
20 and 130 $M_{\odot}$ 
forming massive black holes with mixing--fallback in the ejecta, 
combined with the SN-induced star formation model in which, for a given yield,
 Fe/H is inversely 
proportional to the explosion energy.  Strong mixing and fall-back explains the 
cases with strongly enhanced C and O and corresponds to an observed class of faint 
supernovae (IIp), while higher explosion energies (possibly assisted by rotation) 
produce hypernovae (some of classes Ic and IIn), in which complete 
Si burning leads to $\alpha$-element excess and the peculiarities in the iron group.     

The {\em r}-process, reviewed by John Cowan, still holds many mysteries, since its 
path involves highly unstable nuclei and there is no certainty as to where it occurs;  
Grant Mathews favours the neutrino-heated hot supernova bubble, but neutron star pairs 
are another candidate and in any case magnetic fields, jets and neutrino oscillations 
may be involved. Some intriguing nuclear information has come from H-bomb 
tests (Stephen Becker). 

Among the lowest-metallicity stars, there are huge variations in the relative 
abundance of {\em r}-process and iron group, while among the {\em r}-process 
nuclides themselves 
there is sometimes excellent agreement with the solar-system {\em r}-process distribution 
and sometimes not, expecially between the $A\simeq 130$ Xe peak and the higher 
Pt peak.  So, as previously suggested by Wasserburg et al.\ (1996) on 
other grounds, there may be two or 
more distinct {\em r}-processes, analogous to the weak and main {\em s}-processes. 
This in turn relates to radioactive cosmochronology, to which we return later.  

\subsection{Low and intermediate-mass stars} 
Intermediate-mass stars are significant sources of He, C, N and the main {\em s}-process. 
We heard from Dick Henry that, compared with the old work of Renzini and Voli, 
widely used despite their own health warnings for so many years, more recent 
synthetic models like those of van den Hoek \& Groenevegen predict higher mass loss 
rates and less helium production and an increase in C and N production at lower 
metallicities. Hot-bottom burning sets in at around $3M_{\odot}$ and C and N 
production is further 
enhanced by rotation. There is fair agreement with abundances found in planetary 
nebulae. A somewhat contentious issue is raised by the bimodal distribution of 
N/$\alpha$ ratios in damped Ly-$\alpha$ systems, the upper branch corresponding to 
N/O in H {\sc ii} galaxies and the lower branch nearly an order of magnitude lower; 
could this be an age effect, or is there something funny going on with the initial 
mass function? Mercedes Molla described galactic chemical evolution models 
with a new set of yields for carbon and nitrogen, and Amanda Karakas discussed the 
production of aluminium (including $^{26}$Al) and heavy magnesium isotopes in AGB stars. 

Rotation also influences mixing processes following the first dredge-up  
(Corinne Charbonnel). At the bump in the luminosity function where the H-burning 
shell crosses the previous lowest boundary of the outer convection zone, the 
$\mu$-barrier is partially lifted enabling various extra-mixing processes to take place
as a result of rotation. 
Consequences include lowering of the $^{12}$C/$^{13}$C ratio, destruction of 
$^ 3$He and the production of a lithium `flash' leading to enhanced mass loss 
attested by a dust shell after a brief super-lithium rich phase.  

Significant, perhaps even dominant, contributions to galactic enrichment in $^ 7$Li, 
$^{13}$C, $^{15}$N and $^{26}$Al come from novae (Sumner Starrfield); UV spectra of 
some fast novae show enhanced abundances of N, O, Ne, Mg and Al, but not C or Si.   

Dramatic advances in the theory of the {\em s}-process have resulted from the efforts 
of the Torino group, presented by Maurizio Busso and Oscar Straniero. Evidence from 
branchings in the Kr--Rb region confirms that most of the main {\em s}-process with 
neutrons from $^{13}$C takes place during shell H-burning phases between thermal 
pulses, with just a minor top-up from $^{22}$Ne during the pulses themselves. This 
renders obsolete the old idea of an exponential distribution of exposures 
(Seeger, Fowler \& Clayton 1965), which I described in my book as arguably the most 
elegant result in 
the whole of nucleosynthesis theory! The steps at magic numbers are still there, 
however.  A free parameter in the theory is the mass of the $^{13}$C pocket that 
gets ingested into the intershell zone. This seems to be more or less constant, 
leading to increasing neutron exposures with lowering metallicity, and  
eventually to a `strong' {\em s}-process with lots of lead (confirmed by Judy Cohen).     

\section{Modelling galactic chemical evolution} 

\begin{table}
\caption{{\large  CNO abundances in local group galaxies}}
\label{table1}
\begin{tabular}{|r|c|c|c|l|}
\hline \hline 
&&&&\\
Object                 & C &   N &   O & Reference \\ \hline
\underline{Local Galaxy:} &&&&\\
Sun                   & 8.4  &(7.8?) & 8.7 & All. Pr.\ et al 01,02\\
 "                    & 8.6  & 7.9 &8.7 & Holweger 01 \\
Orion nebula           &8.5 & 7.8 & 8.7 & Esteban et al 98\\
diffuse ISM           &     &     &8.6 to 8.7 & Meyer et al 98\\
cepheids            &8.0/8.6 &&8.3/8.9&Luck et al 98\\  \hline
\underline{M 31:}$\;\;\;\;\;\;\;\;\;\;\;\;$&&&&\\
H {\sc ii} reg.    &  &7.0/8.2  & 8.5/9.2 & Dennef. \& Kunth 81\\
SNR                &  &7.4/8.0  & 8.3/8.7 & Blair et al 82     \\
4 AF supergi     & 8.3 & & 8.8 & Venn et al.\ 00\\ \hline
\underline{M 33:}$\;\;\;\;\;\;\;\;\;\;\;\;\;$&&&&\\
H {\sc ii} reg.    & & 7.9$-.16R$&9.0$-.12R$&V\'{\i}lchez et al 88\\
SNR                & & 7.8$-.12R$& 8.8$-.07R$& Blair et al 85\\
B,A supergi&  & & 9.0$-.16R$ & Monteverde et al\ 97 \\ \hline
\underline{LMC:}$\;\;\;\;\;\;\;\;\;\;\;\;\;$ &&&&\\
H  {\sc ii} reg., SNR & 7.9 & 6.9 & 8.4 & Garnett 99 \\
cepheids          & 7.7/8.3 & & 8.0/8.8 & Luck et al 98\\
{\small PS 34-16 (early B)} &    7.1 & 7.5 & 8.4 & Rolleston\\
{\small LH 104-24 ("")}$\;\;\;$ &   7.5 & 7.7 & 8.5 & et al 96 \\
{\small NGC 1818/D1}    &        7.8 & 7.4 & 8.5 & Korn et al 02\\
{\small N2004 (4 early B)} &  8.1 & 7.0 & 8.4 &$\;\;$   "      "\\
4 F supergi           & 8.1 &     &     & Russell \& Bessell 89 \\ \hline
\underline{SMC:}$\;\;\;\;\;\;\;\;\;\;\;\;\;$ &&&& \\
H  {\sc ii} reg., SNR & 7.5 & 6.6 & 8.1 & Kurt et al 99\\
cepheids &         7.4/7.8 && 8.0/8.3 & Luck et al 98\\
10 A supergi&$<7.3$/$<8.7$&6.8/7.7&8.1& Venn 99\\
3 F supergi            &7.7      &      &8.1&Spite et al 89\\
2 F supergi            &7.8    &     &&  Russell \& Bessell 89 \\ \hline
\underline{NGC 6822:}$\;\;\;\;\;$ &&&&\\
H  {\sc ii} reg. &      &       6.6  & 8.3& Pagel et al 80\\
"  "$\;\;$            &       &            & 8.4& Pilyugin 01\\
2 A supergi &      &         & 8.4 & Venn et al 01\\  \hline
\underline{WLM}:$\;\;\;\;\;\;\;$&&&&\\ 
H  {\sc ii} reg. &      & 6.5       & 7.8 & Skillman et al 89\\ 
A supergi &   & 7.5 &  8.4& Venn 03\\ \hline  
\end{tabular}
\end{table}

According to taste, there is a great variety in the degree of sophistication and 
complexity that one can put into galactic chemical evolution models, ranging from 
chemodynamical and cosmological SPH plus semi-analytic simulations (as described by 
Brad Gibson) to numerical models with parameterized star formation and gas flow laws 
(Francesca Matteucci) to highly simplified analytical toy models, which still have a 
useful contribution to make, as was shown at this meeting by Verne Smith 
and Matt Shetrone, for 
example.   The wide range in {\em r}-process/Fe ratios in the most metal-poor stars 
suggests some kind of inhomogeneous model (Sally Oey), although the simplest version 
of this postulates a threshold that is not observed and predicts far too many 
metal-free stars. 

\section{Stellar abundance analysis} 

Substantial revisions in the `official' solar oxygen abundance in the last two years 
lead to some doubts as to how good stellar abundances really are and the influence 
of bandwagon effects.  Bengt Gustafsson described the activities of `sheep' 
(who follow the crowd) and `goats' (who like to put a spanner in the works, but 
not too big a one). I believe the comparison with emission lines from H {\sc ii} 
regions is a good check (see Table 1.1). To one place of 
decimals in the log (to quote two would be like a second marriage --- a triumph of hope 
over experience!), the agreement is gratifyingly good, or rather was, until Kim 
Venn dropped her WLM bombshell at this meeting! The H {\sc ii} O-abundance is 
typical for such a dwarf galaxy, whereas the stellar one is surprisingly high. 
We heard from Andreas Korn and Norbert Przybilla how important it still is to get 
better atomic data for both LTE and non-LTE abundance analyses, and from Mariagrazia 
Franchini about the possibilities of using the Lick indices to get $\alpha$/Fe 
ratios. 

\section{Abundance effects from internal stellar evolution} 

Dave Lambert discussed the wide range of effects observed in AGB and post-AGB 
stars.  Hot-bottom burning may break through into the Ne-Na and Mg-Al cycles, 
leading to peculiarities in $^{26}$Mg/$^{24}$Mg ratios in NGC 6752 and some 
field dwarfs.  HBB can also  
lead to $^ 7$Li and fluorine production, and a final shell-flash seems to 
be responsible for the effects found in FG Sge and H-poor carbon stars like 
Sakurai's object and R Cr B.  RV Tau stars, a class of post-AGB, are apparently 
metal-poor because refractories are locked on dust, resulting in an enhancement 
of {\em s}-process/Fe ratios which in turn correlate with heavy/light {\em s}-process 
ratios (Maarten Reyniers). There is now direct evidence for {\em s}-process 
enhancements in planetary nebulae (Harriet Dinerstein). 

Light elements are affected by mixing processes even in the main-sequence stage. 
Anne Boesgaard has discovered a beryllium dip in open clusters like the Hyades, 
which shows interesting contrasts to the lithium dip.  Within the dip, Be is down, 
though not as much as Li, which is reminiscent of effects of rotational mixing  
on $^ 6$Li/$^ 7$Li (Pinsonneault et al.\ 1999), whereas on the cool side Be is 
undepleted for quite a long way, which corresponds more to a vertically stratified 
model with a deeper exclusion zone for Be than for Li. 

\section{Abundances in stellar populations} 

\subsection{The Milky Way halo} 

The study of the oldest stars in our Galaxy, some of which appear to have lower 
metallicity than anything seen at high red-shifts up to now, was justly described 
by John Norris as `near-field cosmology'. The records are held by G77--61  
([Fe/H] $\simeq -5.5$) and Christlieb's star ([Fe/H] $\simeq -5.3$), discovered 
in the Hamburg Objective Prism survey, but the numbers fall 
below expectations from 
the Simple model for the halo below [Fe/H] $=-4$ (Norbert Christlieb searched 5 million 
stars to get 2000 extremely metal-poor candidates), a result explained by Tsujimoto et 
al.\ (2000) in terms of SN-induced star formation. Further candidates are likely to 
be found from the huge data base provided by the Sloan Digital Sky Survey with the 
aid of automated spectrum analysis (Carlos 
Allende Prieto).  Lithium abundances on the Spite 
plateau are a bit uncomfortably low for Big Bang nucleosynthesis in the light of 
deuterium and MWB fluctuation spectrum data, so there has to be quite a bit of 
depletion, by a factor of 3 or so, and the presence or absence of a metallicity 
dependence hinges on subtleties in the abundance analysis. Some globular cluster 
stars show  higher lithium abundances and a scatter (Boesgaard).  

Below [Fe/H] $=-3$, very large relative overabundances of the CNO elements are 
sometimes found (see Nomoto's and Judy Cohen's talks), and in a few cases this 
also applies to Mg, Si and Eu, 
but for most stars $\alpha$/Fe is quite flat down to [Fe/H] $=-4$, while Co/Fe 
rises and Mn, Cr/Fe fall with decreasing metallicity.  Among neutron-capture 
elements, Sr varies wildly, in contrast to Ba, and the ratio of {\em r}-process to iron 
is highly variable. 

New data from UVES on the ESO VLT exhibit a remarkably uniform and flat trend 
in $\alpha$/Fe ratios between [Fe/H] $=-2$ and $-4$ (Monique Spite).  I remarked 
that only analytical toy models could account for such a flat trend, but Francesca 
disagreed. According to Monique's data, oxygen shows only a modest overabundance 
down to [Fe/H] $=-3.5$, apart from the odd exception; this hot issue, also 
discussed by Suchitra Balachandran and in a poster by Garik Israelian, was delicately 
tiptoed around by speakers and audience.  

With the discovery of uranium in CS 31082$-001$, radio-active cosmochronology 
has at last 
become respectable (Roger Cayrel).  For a long time, the Th/Eu ratio has been used 
for this purpose, following a suggestion of mine (Pagel 1989), and at first I 
was a bit miffed at this not having been mentioned, but now that this ratio has 
been exposed as a quite unreliable one (see also the poster by Otsuki,  
Mathews \& Kajino 2002), I suppose I really should be grateful!    		

\subsection{The Thin and Thick Disks} 

\begin{figure} 
\centering
\includegraphics[width=7cm, angle=179]{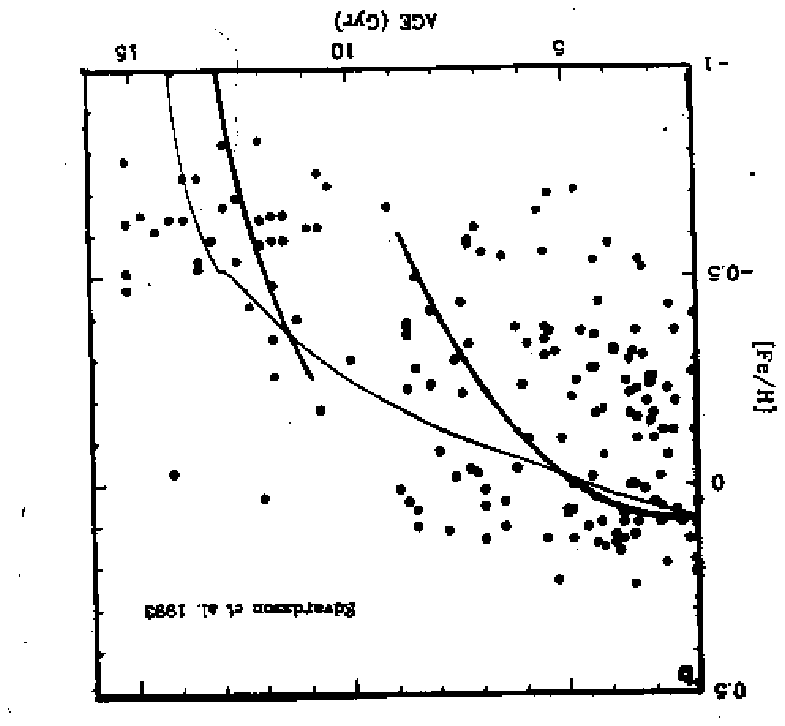}
\caption{Thin curves: age-metallicity relation from the `two-inflow' model of 
Chiappini, Matteucci \& Gratton (1997). Thick curves: sketch of suggested AMRs 
for the Thick and Thin disks.  Data points from EAGLNT (Edvardsson et al.\ 1993). 
Adapted from Chiappini et al.\ (1997), after Pagel (2001).} 
\end{figure}
  
Thick-disk stars (essentially the same as Bengt Str\" omgren's `Intermediate 
Population II') resemble Bulge stars with minor differences, i.e. they contnue 
to display the $\alpha$-rich effect noted by George Wallerstein 40 years ago 
(Wallerstein 1962) 
up to quite high metallicities, symptomatic of an old get-rich-quick stellar 
population, while in the halo both $\alpha$-rich and non-$\alpha$-rich stars 
are found, the latter belonging to the outer halo and showing similarities in composition 
to dwarf spheroidals and irregulars (Poul Nissen). In the Thick Disk,   
however, as appears from recent work reported by Sophia 
Feltzing (as well as the EAGLNT survey and other studies), the $\alpha$/Fe ratio 
does actually come down at high metallicities, indicating a 
significant contribution from SNIa as well as possible effects of a star formation 
threshold (Cristina Chiappini). While the time needed for a significant 
contribution from SNIa is often assumed to be of the order of 1 Gyr, in reality 
there must be a spread, so that it is still hard to be quantitative about 
the age and formation time-scale of the Thick disk, but a sketch of how the 
chemical evolution of both disks might have gone 
is shown in Figure 1. 
A new version of the EAGLNT survey is under way (Bacham Reddy).    

\subsection{The Galactic Bulge} 

Our host Andy McWilliam has been making heroic efforts over the years, in 
collaboration with Mike Rich, to gain accurate abundances for red giants in the 
Bulge.  Their latest results on K-giants show trends in $\alpha$-elements that are 
rather like 
those in the Thick disk, with O, Mg and Si high (as well as Al and Eu), but O/Fe 
coming down at the highest metallicities. Ca/Fe behaves like in the thin disk, 
which is interesting because Lambert et al.\ (1996) found a similar effect in the 
RR Lyrae 
stars, presumably representing the Thick disk. What is this telling us about the 
origin of calcium? Near infra-red spectra enable the abundance analysis to be 
extended to M-giants in globular clusters (Livia Origlia).   

\subsection{Globular clusters} 
 
The GC metallicity scale has been refined by using Fe {\sc ii} (Bob Kraft).  
Galactic globular clusters all have [Fe/H] $\geq -2.4$, but in many respects show 
similar abundance patterns to field stars in the same metallicity range (Chris 
Sneden). The main differences are the Na,Al--O anticorrelation (although Al 
abundances are still a bit dodgy) and the persistence of CNO anomalies down to 
the main sequence, suggesting a multi-generational aspect to globular clusters. 
Ruprecht 106 is exceptional and may be a captured system, and the inner-halo--outer 
halo effects found in field stars have a counterpart in globular clusters, 
e.g. M4 versus M5 (Jon Fulbright).  Still stronger anomalies are found in a few 
cases (Inese Ivans). 

The most intriguing globular cluster (or galaxy remnant?) is, of course, $\omega$ 
Centauri, with metallicities ranging from [Fe/H] $=-2$ to zero, and high {\em s}-process 
abundances (Verne Smith; Elena Pancino). Lanthanum, a heavy {\em s}-process product, 
shows a stepwise increase with Fe/H at [Fe/H] $=-1.6$, while the light {\em s}-process 
element yttrium varies less. $\omega$ Cen 
displays both cluster-like (Na--O anticorrelation) and dwarf galaxy-like properties  
(low $\alpha$/Fe, Eu/Fe). The metal-poor component shows rotation; the metal-rich one 
does not, all of which suggests a merger of two extragalactic globular clusters 
and capture into the Milky Way. A model by Takuji Tsujimoto envisages three star 
formation episodes, the first terminated by a wind whereafter AGB stars enriched 
the remaining gas with heavy {\em s}-process; in the third phase SNIa-induced 
star formation 
occurred and the remaining gas was stripped by passage through the Milky Way disk.     
He ascribes the difference from other GCs to formation by colliding proto-cluster 
clouds; low impact velocities favour SN-induced star formation and chemical evolution, 
whereas high velocities prevent it. 

\subsection{Nearby galaxies} 

The LMC has stars of all ages, despite the gap in the cluster age distribution, and 
a fairly well-defined age-metallicity relation; it remains to be clarified how 
`bursty' the star formation history has been (Vanessa Hill). Carbon and nitrogen 
abundances are low, as in blue compact galaxies, suggesting youth, but $\alpha$/Fe 
and O/Fe ratios are also low, suggesting a long star-formation time-scale. There 
are also substantial differences between clusters of similar age and metallicity in 
both Clouds (Jennifer Johnson). 

Stellar-wind analysis techniques (Fabio Bresolin) applied to A-supergiants in 
gas-rich dwarf galaxies NGC 6822, 
Sextans and GR 8 give oxygen abundances in good agreement with H {\sc ii} regions 
in the same galaxies, and also metal abundances, which indicate [$\alpha$/Fe] $\simeq 
0$ down to [Fe/H] $\simeq -1.5$ (Kim Venn), but for WLM there is a discrepancy 
(see Table 1.1). The common finding of solar 
$\alpha$/Fe is somewhat puzzling: is there something fundamental about it (e.g.\  
as a consequence of star-formation bursts), or does it 
imply largely common star formation histories? These differ considerably in detail, 
although age-metallicity relations are often similar within the uncertainties  
(Eva Grebel).  
In any case, there is not the large variety in $\alpha$/Fe ratios suggested by 
Gilmore \& Wyse (1991).  

High-resolution spectroscopy with large telescopes has also led to detailed 
information about ages and compositions in dwarf spheroidal galaxies (Matt Shetrone). 
These divide into those with a simple star formation history dominated by an early 
burst, e.g. Dra, Sex, UMi, Scl, and those with a more complicated one, e.g.\ Sgr, 
For, Car, Leo I and II. The simple group provides new insights into nucleosynthesis: 
O,Mg/Fe start to go down when [Fe/H] $\geq -1.5$, suggesting that star formation 
was interrupted, and Ca and Ti are less enhanced than O and Mg, due to explosive versus 
hydrostatic synthesis or a contribution from SNIa? The metallicity dependence of 
Mn and Cu is not due to the latter effect, as can be deduced also from their 
behaviour in the Galactic halo. Y and Na are low, while Ba/Y is high, indicating 
{\em s}-process production in metal-poor AGB stars.  

Complex star formation histories place constraints on hierarchical galaxy 
formation models 
(Tammy Smecker-Hane).  Dwarf spheroidals mostly follow a metallicity-luminosity relation, 
$Z\propto L^{0.3}$, as do gas-rich dwarfs though with a lower zero point, 
but not for the reasons given by Dekel \& Silk (1986) which require a 
simple history culminating in a terminal wind, and it seems that total luminosity 
is more important for metallicity 
than are the details of SF history (Carmen Gallart). Furthermore, giant 
elliptical galaxies cannot have been made up from dwarf spheroidals. The main 
difference between gas-rich and gas-poor dwarf galaxies is that the latter have been 
robbed of their gas by the Milky Way and M 31, probably through ram pressure 
(Jay Gallagher). 

The metallicity-luminosity relation extends all the way up to M 31, which has a 
disturbed, metal-rich halo with [Fe/H] $\simeq -0.6$ like 47 Tuc (Mike Rich) and 
many blue horizontal-branch stars indicating great age. The globular 
cluster system rotates and includes a GC/cap\-tured dSph very like $\omega$ Cen.      

\section{The interstellar medium} 

X-ray satellites ASCA and Chandra now permit spatially resolved abundance determinations 
in supernova remnants (John Hughes). In SNIIs these are lumpy and diverse, 
and the accuracy 
of relative abundances depends on the species compared being in the same place. 
Cas A has at least 4 components, of which one is a featureless non-thermal continuum 
while others represent O-burning (Si, S) and incomplete Si-burning (Fe). One can 
have `inside-out' configurations in which iron-group elements overtake the lighter 
ones.  SNIa remnants are smoother, but may contain lumps of hot iron!   

Various solid pieces of stars are found as pre-solar grains in meteorites (Don 
Clayton) and provide constraints on the chemical evolution of the Galaxy. `Mainstream' 
SiC grains come from AGB stars (with relatively low $^{12}$C/$^{13}$C and high 
$^{14}$N/$^{15}$N), but there are also X-grains from supernovae where these ratios 
are reversed. The silicon isotopes present something of a mystery because the plot 
of $^{29}$Si excess vs.\ $^{30}$Si excess has a slope of 1.3 and passes above the 
(solar) origin although the stellar sources must have been older and presumably 
less metal-rich than the Sun.  My suggestion is that there may be a bias towards high 
metallicity in the sample, e.g. if stellar winds are metallicity-dependent. 
Isotopic and elemental patterns in the grains suggest a well-mixed Galactic 
chemical evolution (Larry Nittler). 

However, interstellar dust does not consist of unmodified solid ejecta from stars; 
these are subject to shocks, sputtering etc.\ in the ISM and some return to the gas 
phase with a typical turnover time of $3\times 10^ 8$ yrs and there is grain 
growth in the ISM (Bruce Draine).  Various clues suggest the composition and 
size distribution of the dust: 2200 \AA \ absorption comes from sp$^ 2$-bonded 
carbon in sheets (graphite or PAH), diffuse interstellar bands from large molecules 
(?), 3.4 $\mu$m features from C--H stretches in linear chains and mid-IR features 
from PAHs and Si--O stretches. The ratio of visual absorption to reddening increases 
with the size of the particles, which is mostly under 1$\mu$m.  

UV spectra (from Copernicus to FUSE) reveal how much of the standard abundance 
distribution is depleted from the gas to the dust phase, but sometimes ionization 
corrections are needed (Ed Jenkins). The revised solar O and C abundances lead to 
a more consistent picture than one had before (cf.\ Table 1.1), and there is now 
some information 
about the composition of high-velocity clouds and the Magellanic Stream, which 
resembles that in dwarf Irregular/blue compact galaxies. In the Galactic plane, 
the D/H ratio is uniform in the local bubble (100 pc), but shows anomalous variations 
further afield -- only $7.5\times 10^{-6}$ on the sight-lines to $\lambda$ Sco 
and $\delta$ Ori, but $2.2\times 10^{-5}$ on that to $\gamma^ 2$ Vel (Jeff Linsky). 
Could deuterium be locked on grains? This would hardly account for the unusually 
large value, but a spatially and temporally varying infall of relatively unprocessed 
matter might. No deuterated molecules were detected in a cloud 28 kpc from the 
Galactic centre by Don Lubovitch, but molecular features are rather weak there 
anyway.        

Absorption lines of molecules such as CO, CN, HNC and HCO$^ +$ can now be studied by 
a new technique using interferometry (Tommy Wiklind). In both diffuse and dark 
clouds, CO and HCO$^ +$ are much more abundant than expected from gas-phase 
reaction networks, while O$_ 2$ is expected but not seen. CO has been detected in 
emission from FIR luminous galaxies/AGNs up to a red-shift of 4.7 and many molecules 
have been found in absorption up to $z=0.9$; relative abundances are similar to 
those found locally. 

\section{The local universe}  

Galactic winds resulting from supernova activity in starburst galaxies can 
strongly influence chemical evolution, depending on the depth of the potential 
well and the structure of ambient gas (Crystal Martin). 
Hot SN ejecta are removed in the wind, but not much of the ISM. High-density 
winds appear in X-ray images (e.g. NGC 3077 and NGC 1569), whereas low-density 
winds are detected from blue-shifted absorption lines, e.g. in ULIRGS. Mass 
flow rates are comparable to star formation rates, but in big galaxies it is 
not clear that all the material involved escapes, nor is it clear whether 
large or small galaxies make the dominant contribution to the intergalactic medium. 

Abundances in stellar populations can be studied on the basis of integrated light 
using the Lick indices (Scott Trager) calibrated on globular clusters. The 
age-metallicity degeneracy is broken using H$\beta$, provided there is no emission 
or extended blue horizontal branch, but one does need accurate and complete 
isochrones and a spectral library. Some results indicate nitrogen enhancements 
in red giants of M 31 and Fornax, but not in old clusters of the LMC, while 
conversely these have an $\alpha$/Fe enhancement while Fornax does not. 
Ellipticals in the Coma cluster are older than field ellipticals, both with 
$\alpha$ enhancement, while S0s have a spread in age with less $\alpha$ 
enhancement, but the ages are less robust than the chemical results, which will 
be extended to less prominent elements in the near future. The calcium triplet 
in ellipticals is anomalously weak, either because of bad fitting functions or 
because of some anomaly in Ca/Mg (see Galactic Bulge!).    

Another approach to composition variations between and across galaxies comes 
from H {\sc ii} regions (Don Garnett). There are some problems, including 
temperature fluctuations and the calibration of $R_{23}$ (which I sometimes 
think of like Macbeth as `Bloody instructions which, being taught, return to plague the 
inventor'!), but there is quite good agreement with young stars (Table 1.1). 
Oxygen abundance gradients in non-barred spirals such as M 101 are quite constant 
at $-0.2$ dex 
per scale length, a simple result that may need a complicated explanation.  
The metallicity-luminosity relation translates into an effective yield 
(abundance/ln gas fraction) which increases with rotational velocity (pp mass) 
up to a point and then levels off, presumably when SN-driven winds no longer 
escape. Iron abundances can sometimes be found from Fe {\sc ii} lines --- 
prominent in AGNs and some peculiar stars like $\eta$ Carinae ---  taking account 
of UV pumping from Ly-$\alpha$ and other sources, e.g. in Orion it is about 
1/10 solar in the gas phase (Ekaterina Verner). 

One of the benefits of the SDSS is the possibility to study 
the star formation rate density in the local universe from a complete sample 
of H$\alpha$ emission (Jarle 
Brinchmann). The major contribution comes from dusty high-metallicity 
high surface brightness spirals 
and the SFR is about 1/4 of the past average, in agreement with the results of 
Madau et al. 

ASCA and XMM-Newton spectra of numerous clusters of galaxies have provided new 
details about the composition of the intra-cluster medium (Michael Loewenstein).  
There is no evolution 
with red-shift up to $z=0.8$, but some variation with temperature of the X-ray gas. 
CNO/Fe are as in the Galactic disk, with subsolar metallicity, but Si and Ni are 
relatively overabundant and the composition cannot be represented by any 
combination of conventional SNIa and SNII. Could there be a Population III contribution? 

\section{The high red-shift universe} 

The composition of the broad emission-line gas in quasars bears witness to the 
effect of rapid star formation accompanying that of the central black hole 
(Fred Hamann). In other words, a get-rich-quick population suitable for the BH's 
future role as the core of a massive spheroidal system.  Absolute abundances are 
model-dependent, but the high relative 
abundance of N compared to C and especially O is suggestive of a high metallicity 
like twice solar, and the mass of the region is similar to that of a globular 
cluster.  Intrinsic, narrow absorption lines confirm at least solar abundance of 
carbon. 

Next in metallicity (and in ambient mass density) after quasars come the Lyman 
break galaxies (Kurt Adelberger), about which much has been found out from the 
lensed object cB~58.  The abundance of oxygen is about 1/3 solar, while the 
interstellar lines indicate a lower abundance of iron-group elements in the 
gas phase. The P Cygni profile of C {\sc iv} resembles that of stars in the LMC. 
Ly-$\alpha$ has a violet-shifted component indicating outflows at 300--600 km s$^{-1}$, 
which perturb and enrich the intergalactic medium out to 0.5 Mpc, leading to a 
correlation between LBGs and intergalactic C {\sc iv}. So we witness supernova 
feedback in action. 

Next in the scale come the damped Lyman-$\alpha$ systems which may be the raw 
material for disk galaxies today (Jason Prochaska, Paolo Molaro, Francesco 
Calura). Metallicity, measured by 
zinc, shows only a mild evolution with red-shift from 2 to 6 and there are problems 
from dust: selection bias and differential depletion from the gas phase. As was 
already mentioned by Dick Henry, N/$\alpha$ is bimodal, the low branch being 
attributed by Jason to massive Pop III stars, but I prefer Paolo's explanation in 
terms of the yield from conventional massive stars, enhanced by rotation as described 
by Meynet \& Maeder (2002).  Sub-DLA systems, with N(H{\sc i}) $\simeq 
10^{19}$ cm$^{-2}$, contribute significantly to the gas and metal budgets and 
show stronger evolution with $z$ (C\' eline Peroux). 

Finally we come to the intergalactic medium, aka the Lyman forest (Bob Carswell, 
Rob Simcoe), which is sparse at low red-shift and best seen at $z=2$ to 3. Usable 
lines are from C {\sc iv}, N {\sc v}, Si {\sc iv}  and (buried in the Ly-$\alpha$ 
forest) O {\sc vi}.  O {\sc vi}/ C {\sc iv} gives electron densities agreeing with 
SPH simulations, and this with ionizing flux estimated from Si {\sc iv}/ C {\sc iv}
enables some abundance estimates to be made: [C/H] $\simeq -2$ in the neighbourhood 
of galaxies. More sensitivity is obtained by stacking  C {\sc iv} spectra and 
computing pixel optical depths; some weird enriched regions have been found in this 
way. No chemically pristine regions have been found (Rob Simcoe), typically 
[C/H] $\simeq -2.5 $ whenever Ly-$\alpha$ is seen, although it can go down as low 
as $-3.5$; the problem is that one is fighting against ever diminishing column 
densities!      

\section{Conclusions}

\begin{table} 
{\large  
\caption{Baryon and metal budgets, after Finoguenov et al.\ (2003)} 
\begin{tabular}{llllll}
\hline \hline 
\multicolumn{6}{c}{$z=0$} \\ \hline  
Component& $Z, 10^{-2}$& $\Omega_{\rm Z},\;10^{-5}$&& $\Omega_{\rm b},\;10^{-2}$ \\ 
\hline  
Stars& 1.2\footnote{i.e. solar.} &2.3 -- 4.6& --- Most& 0.2 -- 0.4& \\    
O {\sc vi} absorbers & 0.26 &1.8 -- 5.2& --- metals & 0.7 -- 2.0 & --- Most\\ 
Ly-$\alpha$ forest & 0.01 &0.1&&1.2 & --- baryons\\ 
X-r gas, clusters& 0.7& 1.4&&0.2\\ 
Total & 0.28 &5.6 -- 11.3  \footnote{$\Rightarrow$ yield $\equiv\Omega_{\rm Z}/
\Omega_{\rm stars} \simeq 0.028$}& &2.3 -- 3.8\\ 
Predicted& & &&3.9\\ 
\hline 
\multicolumn{6}{c}{$z=2.5$} \\ \hline  
Protocl. gas & 0.7 & 1.4 &  --- Most &0.2 \\  
ISM, dust & 0.8 & 0.8 -- 1.7 & --- metals& 0.1 -- 0.2 & \\ 
Ly-$\alpha$ forest & 0.01 & 0.1 -- 0.3 && 1 -- 5& --- Most baryons\\ 
DLAs & 0.1& 0.1& & 0.1 \\ 
Total& 0.1 & 2.4 -- 3.5 && 1.4 -- 5.5\\
Predicted&&1.6 -- 3.2 && 3.9\\ \hline   
\end{tabular} 
} 
\end{table} 

Highlights of this conference have been in my view the new insights into the 
{\em s} and {\em r}-processes and the explosive increase in details of stellar 
abundances in the Galactic halo and in dwarf spheroidals, together with their 
star formation history and age-metallicity relations. 

I do not think a conference on the origin and evolution of the elements 
would be complete without a survey of where the baryons and metals are in the 
universe as a whole, and where they were at a substantial red-shift --- questions 
that have been addressed in the last few years by Persic \& Salucci (1992) and  
Fukugita, Hogan \& Peebles (1998), as far as baryons are concerned, and including 
metals by Pettini (1999), Pagel (2002) and  Lilly,  Carollo \& Stockton 
(2002). Table 1.2 is 
based on work by Finoguenov, Burkert \& B\" ohringer (2003), 
whose numbers I quote with their kind permission.     

While it has long been known that stars only account for a tenth or so of the 
baryonic matter density deduced from primordial deuterium and the CMB angular 
fluctuation spectrum, there has been uncertainty as to where the missing baryons 
reside at present, although at $z=2.5$ they are likely to be in the ionized gas 
associated with the Lyman forest.  There is probably less such gas around today, 
much of the remainder being associated with O {\sc vi} gas, which also contains 
like half the metals, assuming 0.2 solar abundance (the metal density is somewhat 
more robust than the baryon density, according to Mathur, Weinberg \& Chen 
2003); the other half is mainly in stars.     

Pettini (1999) drew attention to the fact that, while a quarter of all stars 
had been born by a red-shift of 2.5, it was not possible to account for the 
corresponding quarter of today's metals on the basis of DLA's, LBG's or the 
Lyman forest.  This has given rise to two bold, but possible hypotheses 
involving dust in SCUBA galaxies (Dunne, Eales \& Edmunds 2003) and the 
arguments of Finoguenov et al.\ based on non-evolution of intra-cluster gas and 
its metal content since $z=3$, so together these two sites might account for 
the bulk of the missing metals.  Finoguenov et al.\ argue for early enrichment 
of the intra-cluster/proto-cluster gas by a top-heavy IMF. 
I am not sure if this is necessary; the issue may be decidable from the sort 
of data presented by Michael Loewenstein. In any case,  
the numbers in the table indicate an average yield for the whole universe of 
about twice solar, similar to what one can get from a conventional Salpeter 
mass function.

\end{document}